\documentstyle[aps]{revtex}

\begin{document}
\title{In Search of Exact Solutions }

\author{Michel H\'{e}ritier}
\address{Laboratoire de Physique des Solides d'Orsay, Centre Scientifique d'Orsay,
91405 Orsay, France.}
\maketitle
\bigskip
\begin{abstract}
Many-body systems, such as electrons flowing in a superconductor, are among
the most difficult theoretical problems to study. A new family of exactly
solvable models may offer some answers.
\end{abstract}
\bigskip
\bigskip
During the past 40 years, understanding quantum systems involving
many particles has been one of the main goals of theoretical
physics. This kind of research is designed not to work out the
laws of interactions between particles, but, assuming that these
interactions are already known, to calculate their effects on a
system of $n$ particles. A system where $n$ is large constitutes
the `many-body problem', and can exhibit rich behaviour, including
phase transitions, superconductivity and Bose-Einstein
condensation. The impact of exactly solvable theoretical models on
research into these systems is undeniable. With few exceptions,
previous exact solutions have applied only to
one-dimensional systems. But writing in Physical Review Letters, Dukelsky $%
et.~al.$ \cite{uno} introduce a new family of exactly solvable
theoretical models dealing with quantum many-body systems in any
number of dimensions: one, two, three or even more.

A wide range of many-body systems has been studied, with varying
numbers of particles. For example, systems of nuclei ($n\approx
10^{2}$ nucleons) are important in nuclear physics, and atomic and
molecular physics deal with systems involving $n\approx 10^{2}$
electrons, whereas mesoscopic ($n\approx 10^{2}-10^{6}$) and
macroscopic ($n\approx 10^{23}$) systems arise in condensed-matter
physics. Real systems are usually extraordinarily complex and
involve a large set of parameters, most of which are irrelevant
for discussing the phenomena under investigation. So theoretical
physicists begin by defining a model  that is, an idealized system
much simpler than the real one, but retaining all the necessary
ingredients to discuss the physical properties of interest.
However, there are only a few examples of `exactly solvable'
models of this type in which one can calculate exactly all the
possible quantum states and their energies and/or the various
thermodynamic quantities of the system. Such exact solutions
always play a decisive role in theoretical physics.

With very few exceptions, the exact solutions of many-body
problems can be found only in one dimension. For example, the
Tomonaga-Luttinger solution for a one-dimensional system of
interacting fermions (particles such as electrons) demonstrates
how the Landau-Fermi liquid description of electrons, which has
been the theoretical basis of the whole quantum theory of solids,
breaks down in one or two dimensions. This discovery had a
considerable impact on condensed-matter physics, and gave rise to
a new field concerned with the properties of the one-dimensional
state known as the `Luttinger liquid'. In contrast, several
exactly solvable models have been developed in nuclear physics
that make use of group-theory techniques. These models are based
on the concept of dynamical symmetries, and a non-trivial
generalization of this approach is the `pairing model', in which
particles that form a pair interact through a force with a
particular strength. Such a model has been applied to discussions
of superfluidity, and also to studies of superconductivity in
small metallic grains.

The pairing model was solved exactly by Richardson \cite{dos} more
than 30 years ago, a result that passed almost unnoticed despite
its enormous potential impact on nuclear as well as
condensed-matter physics. In their paper, Dukelsky and colleagues
\cite{uno} provide an important generalization of Richardson's
work, leading to new classes of pairing models that are exactly
solvable in any dimension. To obtain this result, the authors
apply group-theoretical and algebraic methods similar to those
already used by Gaudin \cite{tres}. They discover three classes of
integrable (a remarkable property that makes them mathematically
tractable) pairing models, which they were able to solve exactly.

In analogy with the work of Gaudin, the three new classes of
pairing models found by Dukelsky $et.~al.$ \cite{uno} are named
the rational, the trigonometric and the hyperbolic models. The
authors go on to apply the rational model to a two-dimensional
system of $18$ fermions with repulsive interactions, for which
they find an exact numerical solution. Their first results
indicate that there are attractive correlations between pairs of
fermions, despite the underlying repulsive interaction. Such
models might apply to high-temperature superconductivity, in which
mutually repulsive electrons have to pair up to flow without
resistance. In further work, Dukelsky $et.~al.$ \cite{cuatro}
propose an exact correspondence between the solution of the
quantum pairing model and a two-dimensional classical
electrostatic problem: the trigonometric model can be mapped to an
electrostatic problem on a spherical surface, whereas the
hyperbolic model seems to correspond to a mapping on a more
general surface.

The new solutions of the pairing models seem to be limited to
systems of finite size  a necessary condition for solving
equations giving exact quantum states. But they can certainly be
applied to systems where $n$ is quite large, from which the
extrapolation to infinite $n$ is fairly accurate. The
solutions are also restricted to systems at a temperature of absolute zero (%
$-273^{\circ}$ C), but because the exact solution provides all the
quantum states and their energies, any thermodynamic quantity can
be calculated at a non-zero temperature by thermal averaging,
according to the Boltzmann formula of statistical mechanics. Exact
solutions, as well as predicting the particular properties of a
simplified system, can also indicate how good a given
approximation is to a many-body system.

In all these respects, the models of Dukelsky $et.~al.$ are highly
welcome. Another useful feature of these models is that they apply
equally to boson and fermion systems. All known particles are
either fermions (such as electrons) or bosons (such as photons),
which obey different quantum statistics. As the authors show
\cite{uno}, the rational model may have some bearing on fermion
systems, in particular the physics of high-temperature
superconductors. In the case of boson systems, Dukelsky and Shuck
\cite{cinco} have applied the rational model to the transition of
a group of low-temperature atoms to a Bose-Einstein condensate, a
state of matter in which all the atoms share the same quantum
state. In similar work, Dukelsky and Pittel \cite{seis} applied
the model to a system of bosons with short-range repulsive
interactions, and obtained new evidence for the validity of the
`interacting boson model' of nuclear physics.

The physics of strongly correlated fermion systems is one area of
active research that should benefit immediately from these new
families of exactly solvable models. Despite considerable
theoretical efforts, mainly motivated by the problems raised by
high-temperature superconductivity, such systems remain a huge
challenge to theorists. These, and other seemingly intractable
problems, would welcome any insight offered by a new family of
theoretical tools. The results obtained by Dukelsky $et.~al.$
\cite{uno} for a two-dimensional system of fermions with repulsive
interactions is a first step in this direction.

\bigskip

Michel H\'{e}ritier is in the Laboratoire de Physique des Solides
d'Orsay, Centre Scientifique d'Orsay, 91405 Orsay, France.

e-mail: heritier@lps.u-psud.fr

\end{document}